\documentclass[prb,superscriptaddress,twocolumn]{revtex4-2}
\usepackage{graphicx}
\usepackage{amsmath}
\usepackage{braket}
\usepackage{color}
\usepackage{float}
\begin{document}
\title{Origin of the transitions inversion in rare-earth vanadates}
\author{Xue-Jing Zhang}
\affiliation{
Peter Gr\"unberg Institute, Forschungszentrum J\"ulich, 
52425 J\"ulich, Germany}
\author{Erik Koch}
\affiliation{
J\"ulich Supercomputing Centre, Forschungszentrum J\"ulich, 
52425 J\"ulich, Germany}
\author{Eva Pavarini}
\affiliation{
Peter Gr\"unberg Institute, Forschungszentrum J\"ulich,
52425 J\"ulich, Germany}
\date{\today}
\begin{abstract}
The surprising inversion of the orbital- and magnetic-order transition temperatures in the $R$VO$_3$ series
with  increasing the rare-earth radius makes the series unique  among orbitally-ordered materials.
Here, augmenting dynamical mean-field theory  with 
a decomposition of the order parameter into irreducible tensors, we show that this anomalous behavior emerges from an unusual hierarchy of interactions. First,  increasing the rare-earth radius,  orbital physics comes to be controlled  by $xz$-$xz$ quadrupolar super-exchange rather than by lattice distortion. Next, for antiferromagnetic spin order, orbital  super-exchange terms with different spin rank compete, so that the dipolar spin-spin interaction dominates. 
Eventually,  G-type magnetic order (anti-ferro in all directions) can  appear already {\em above} the orbital ordering transition,
and C-type  order (anti-ferro in the $ab$ plane) right around it.
The strict constraints we found  explain why the inversion is rare, giving at the same time criteria to look for similar behavior in other materials.
\end{abstract}

\maketitle

{\em Introduction.}
The interplay of the spin, orbital and lattice degrees of freedom in strongly-correlated transition-metal oxides results in a plethora of phenomena of puzzling complexity \cite{Sci2000,Sci2005,KB}.  In recent years,  heterostructuring,  disorder and non-equilibrium techniques \cite{Interface,diso,nonequilibrium} have opened the path to discovering or engineering new phases. 
In this panorama, the series of $t_{2g}^2$ perovskites {$R$VO$_3$ (where $R$ is a rare-earth)   is paradigmatic \cite{Ren2003,Tokura2003,Goodenough2004,Palstra2007,Khaliullin2007,structure,Fernandez-Diaz2003,Ren2007,Goodenough2008,Yan2019,Gruninger2013,Palstra2006,Gatti2021,Parmigiani2012,Sund2023,Meley2021,Nasir2023,Palstra2007,Keimer2011,Sundaresan2017,Palstra2001,Keimer2008}. 
In fact, they exhibit a set of different phases, structural and electronic, whose onset depends on the rare-earth radius $R_I$. They are   orthorhombic GdFeO$_3$-type perovskites at high temperature  and
 become monoclinic  at a temperature $T_{\rm S}$;
the orthorhombic phase is, however, either re-entrant at $T_{\rm S'}{<}T_{\rm S}$ (small $R_I$),   or coexists with the monoclinic phase in a large temperature window (e.g., for intermediate $R_I$) \cite{Palstra2006,Palstra2007}.  
Structural changes have been identified with changes in orbital ordering (OO). %

The truly peculiar aspect of the phase diagram, however, is  the reversal
of the spin- and orbital-ordering transition with increasing $R_I$.
 For small $R_I$, an anti-ferro (AF) magnetic transition occurs at $T_{\rm N}{<} T_{\rm S}$,  of C-type in the monoclinic and G-type in the orthorhombic phase %
\cite{Palstra2001,Tokura2003,Palstra2007,Keimer2011,Keimer2008,Sundaresan2017}. Orbital fluctuations, however,
are already suppressed in the high-temperature phase \cite{Reul,Molly2007,xj2022}, i.e., OO itself arises at a temperature $T_{\rm OO}{>}T_{\rm S}$}. %
This is the classic scenario for OO  -- in fact,  in almost all known orbitally-ordered materials, from LaMnO$_3$ and KCuF$_3$ to rare-earth titanates, OO precedes spin ordering  ($T_{ N}{<}T_{\rm OO}$).
 Yet, in RVO$_3$ systems, increasing $R_I$, magnetic and
orbital transition approach each other ($T_{ N}{\to}T_{\rm OO}$), and eventually invert around Ce and La \cite{Tokura2003,Fernandez-Diaz2003,Ren2007},
a highly unusual behavior. 
The complex phase diagram  of the rare-earth vandates is believed to be the result of the interplay of lattice distortion, leading
to a sizable intra-$t_{2g}$ crystal-field (CF) splitting, and
super-exchange (SE) effects of the  Kugel’ and Khomskii (KK) type \cite{KK}.
Indeed, we recently have identified LaVO$_3$  as an orbitally-ordered system of the KK kind \cite{2xj2022}.
What, however, controls the surprising inversion of the orbital- and magnetic-order transition with increasing $R_I$ is not understood. 

\begin{figure}[b]
\centering  
\includegraphics[width=0.46\textwidth]{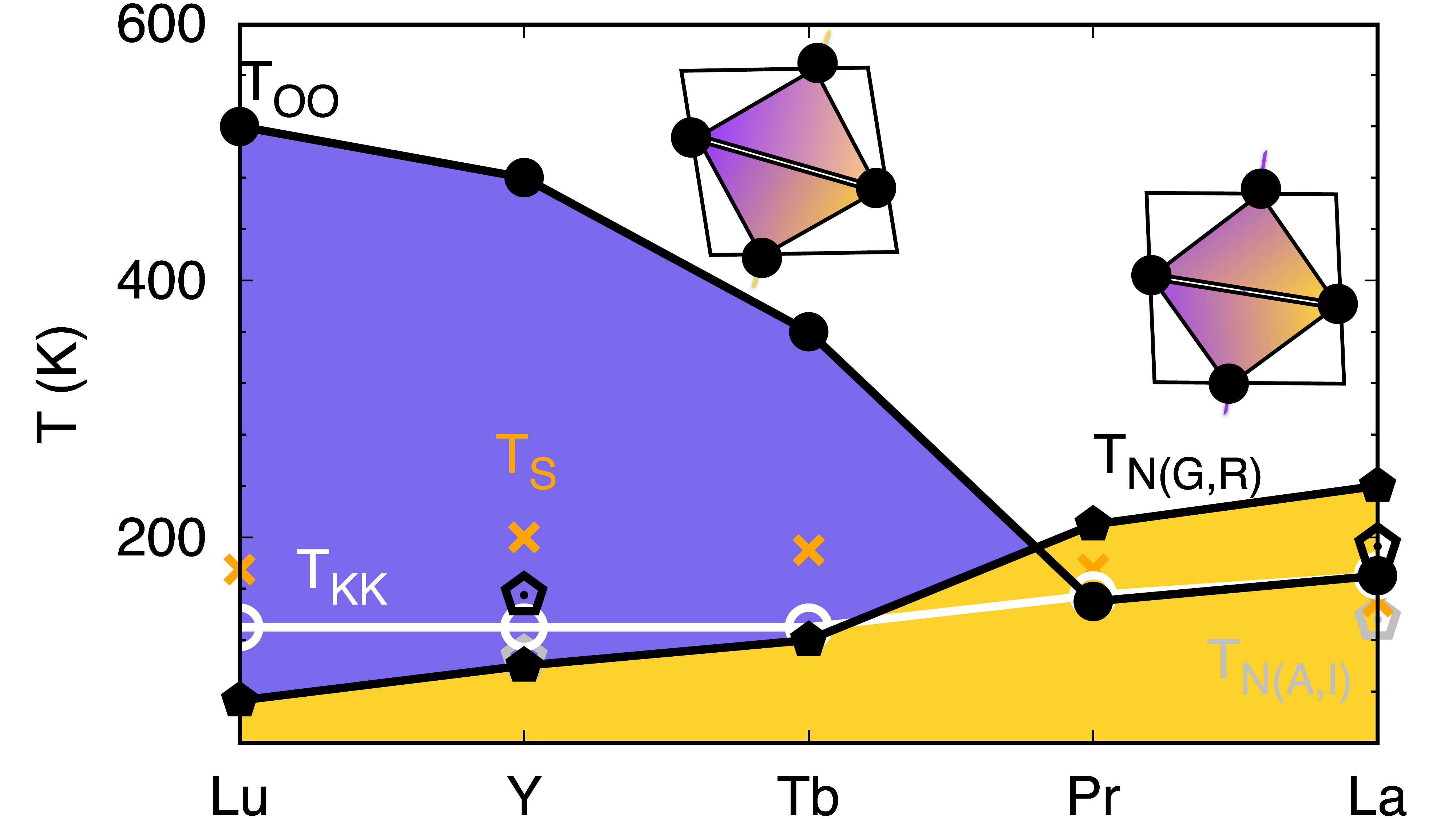}
\caption{
Phase diagram. Filled black circles: orbital-ordering temperature $T_{\rm OO}$. Empty white circles:  Kugel-Khomskii transition temperature $T_{\rm KK}$. Hexagons:  magnetic transition temperature $T_{\rm N}$; 
  G-type (black) and A-type (grey) AF. $R$: real structure.
$I$: idealized case with no CF splitting. $T_S$: experimental structural transition (from \cite{Tokura2003}).
}\label{temp}
\end{figure}

In this work we address and solve this problem. We show that  the inversion is the result of an unusual balance of interactions. 
When $R_I$ is small,  lattice  distortions suppress
the most efficient orbital SE channels. 
This yields the ``classical" picture \cite{oo1,oo2,xj2020,oo4,oo5}
with the orbital state mostly controlled by the CF splitting and $T_{\rm N}{<}T_{\rm OO}$.
Increasing $R_I$, however, reinforces  $xz$-$xz$ quadrupolar SE, which, eventually,
controls orbital physics; 
 yet,  AF spin-spin interactions grow as well, while, at the same time, orbital SE terms of different spin rank compete. 
In the end,  the energy balance tilts, leading to   G-type AF magnetic ordering {\em preceding} OO.
The resulting phase diagram is shown schematically in Fig.~\ref{temp}.

\begin{figure}[t]
\centering  
\includegraphics[width=0.47\textwidth]{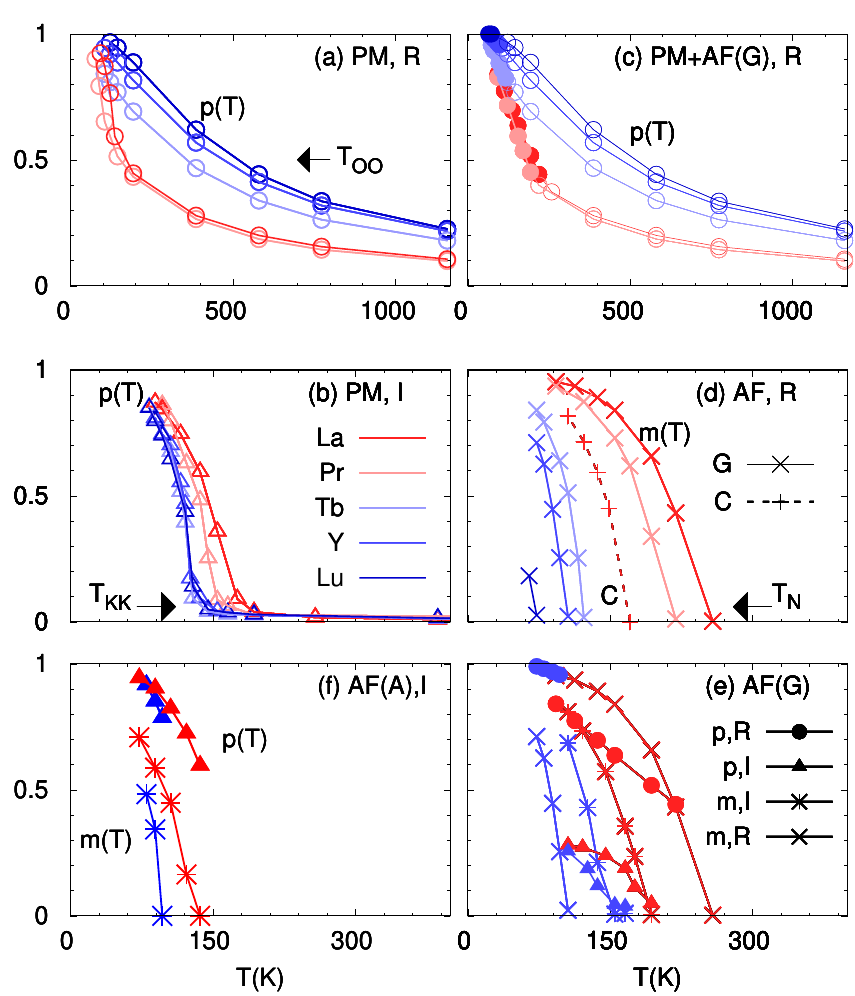}
\caption{LDA+DMFT results. Orbital and magnetic transitions in the $R$VO$_3$ series. Order parameters:  $p(T)$ (orbital polarization) and  $m(T)$ (magnetic moment). Open circles: $p(T)$, paramagnetic (PM) phase. Filled circles: $p(T)$, 
AF phase.  $I$: idealized case with no CF splitting.  $R$: experimental structure; $T_{\rm OO}$: temperature  yielding $p$=0.5. AF: antiferromagnetic 
of G-, C-, A-type.
}
\label{dmft}
\end{figure}

{\em Model and Method.}
We adopt the local-density approximation+dynamical mean-field theory (LDA+DMFT) approach.
First we perform local-density-approximation calculations using the full-potential linearized augmented plane-wave method as implemented in the WIEN2K code \cite{BLAHA1990399}. Next we construct  Wannier functions 
spanning the $t_{2g}$ bands using the maximal localization procedure \cite{PhysRevB.56.12847,wannier90}. Finally, we build the associated Hubbard model 
\begin{align} \nonumber 
\hat{H}=&-\sum_{ii^{\prime}\sigma} \sum_{mm^{\prime}} t^{i,i^{\prime}}_{mm^{\prime}}c^{\dagger}_{im\sigma}c^{\phantom{\dagger}}_{i^{\prime}m^{\prime}\sigma}+U\sum_{im}\hat{n}_{im\uparrow}\hat{n}_{im\downarrow} \\  \label{hub}
&+\frac{1}{2}\sum_{i\sigma \sigma^{\prime}} \sum_{m\neq m^{\prime}}(U{-}2J{-}J\delta_{\sigma,\sigma^{\prime}})\hat{n}_{im\sigma} \hat{n}_{im^{\prime} \sigma^{\prime}}  \\ \nonumber
&-J\!\! \sum_{ i m\neq m^{\prime}}(c^{\dagger}_{im\uparrow} c^{\dagger}_{im\downarrow} c^{\phantom{\dagger}}_{im^{\prime}\uparrow} c^{\phantom{\dagger}}_{im^{\prime}\downarrow}{+}c^{\dagger}_{im\uparrow} c^{\phantom{\dagger}}_{im\downarrow} c^{\dagger}_{im^{\prime}\downarrow} c^{\phantom{\dagger}}_{im^{\prime}\uparrow}).
\end{align} 
Here $t^{i,i^{\prime}}_{mm^{\prime}}$ is the hopping integral from orbital $m$ on site $i$ to orbital $m^{\prime}$ on  site $i^{\prime}$, and
$\varepsilon^{i,i}_{mm^{\prime}}{=}-t^{i,i}_{mm^{\prime}}$ is the crystal-field matrix.
The operator $c^{\dagger}_{im\sigma}$ ($c_{im\sigma}$) creates (annihilates) an electron with spin $\sigma$ in Wannier state $m$ at site $i$, and $n_{im\sigma}{=}c^{\dagger}_{im\sigma}c_{im\sigma}$. The screened Coulomb parameters adopted are $U{=}5$ eV and $J{=}0.68$~eV, established values for this type of systems \cite{PhysRevB54.5368,3d1b,Molly2007}. We solve this model using the dynamical mean-field theory. We adopt the generalized hybridization-expansion continuous-time quantum Monte Carlo method \cite{RevModPhys.83.349}, in the implementation of Refs.~\onlinecite{lamno3c,julian1,julian2}, for the solution of the quantum impurity problem. We define the orbital polarization,  the order parameter for OO, as $p(T){=}(n_3+n_2)/2 {-} n_1$, where $n_i$ are the occupations of the natural orbitals, ordered such that $n_{i+1}\ge n_i$. 
The magnetization  is $m(T){=}(n_\uparrow{-}n_\downarrow)/2$, where $n_{i\sigma}{=}\sum_m n_{im\sigma}$.
In the $t_{2g}^2$ configuration, in the atomic limit, the ground multiplet is $^3P$, a spin and orbital triplet.
The  orbital triplet states can be written as $| \overline{m}_3 \rangle{=}| m_1m_2\rangle$, where  $m_1\ne m_2$ are the occupied orbitals and $\overline{m}_3$ the empty orbital,  with $m_i{\in}span(xy,xz,yz)$ \cite{xj2022}. Using this notation,   the hole orbital at site $i$ is
\begin{align}\label{state}
|\theta,\phi\rangle^i_\alpha %
=&\sin\theta \cos\phi |xz\rangle + \cos\theta |xy\rangle + \sin\theta \sin \phi |yz\rangle.
\end{align}
In the GdFeO$_3$-type structure, there are four (equivalent) V sites in the unit cell. The reference site is $i{=}1$, and  for simplicity we set $|\theta,\phi\rangle^1_\alpha{=}|\theta,\phi\rangle_\alpha$;  the hole-orbitals for the remaining sites are constructed by symmetry \cite{symmetry}. The label $\alpha$ in Eq.~(\ref{state}) specifies how the state is obtained: $|\theta,\phi\rangle_{\rm OO}$, from LDA+DMFT calculations for the experimental structure;  $|\theta,\phi\rangle_{\rm KK}$,  from LDA+DMFT calculations for  an idealized structure with no CF splitting;
$|\theta,\phi\rangle_{\rm M}$, maximizing the SE total energy gain
(see next section).
Finally, the highest energy CF state is  $|\theta,\phi\rangle_{\rm CF}$.

{\em Super-exchange Hamiltonian.}
The crucial element of the puzzle is the materials-specific SE Hamiltonian \cite{KK}. 
Recently we introduced a systematic approach to build it from (\ref{hub})  
 via  irreducible-tensor decomposition \cite{xj2020,xj2022}.  This yields $\hat{H}_{\rm SE}{=}\frac{1}{2}\sum_{ij}  \hat{H}_{\rm SE}^{i,j}$, with
\begin{align}\label{SE}
\hat{H}_{\rm SE}^{i,j}=&\sum_{qq'}\sum_{\nu\nu'}  \sum_{rr'} \sum_{\mu\mu'} \hat{\tau}_i^{r\mu;q\nu} D^{ij;q\nu}_{r\mu,r'\mu'} \hat{\tau}_{j}^{r'\mu'; q\nu} 
\end{align}
where $r{=}0,1,2$ is the orbital rank (monopole, dipole, quadrupole) with components $\mu=-r,\dots,r$, and $q=0,1$
(monopole, dipole) is the spin rank with components $\nu=-q,\dots,q$. It is convenient to split Eq.~(\ref{SE}) as
\begin{align}
\hat{H}_{\rm SE}^{i,j}=\hat{H}_{C_{ij}} + \hat{H}_{O_iO_j} + \hat{H}_{S_iS_j} + \hat{H}_{S_iS_jO_iO_j}.
\end{align}
The first term, $\hat{H}_{C_{ij}}$, obtained by setting $r{=}r'{=}q{=}q'{=}0$, in the absence of charge ordering, is a constant, independent on orbital and spin state; here we take it as the energy zero.
The second term, $\hat{H}_{O_iO_j}$, obtained by setting $q{=}0$ and $r{+}r'\ne0$, describes the interaction between orbital pseudospins,
 independent of the magnetic state. The third term ($r{=}r'{=}0$ and $q{\ne}0$) describes a pure spin-spin interaction, independent of the  orbital state. Finally the last term, obtained
for  $r{+}r'\ne0,q{=}1$, is $\hat{H}_{S_iS_jO_iO_j}$, and describes the entangling of spins and orbitals.
The analytic expressions of the tensor elements $D^{ij;q\nu}_{r\mu,r'\mu'}$ can be found in Ref.~\onlinecite{xj2022}. They depend on $U,J$ and the hopping integrals; the latter are listed in the Supplemental Material \cite{SM}. 
\begin{figure*}[t]
\centering  
\includegraphics[width=0.95\textwidth]{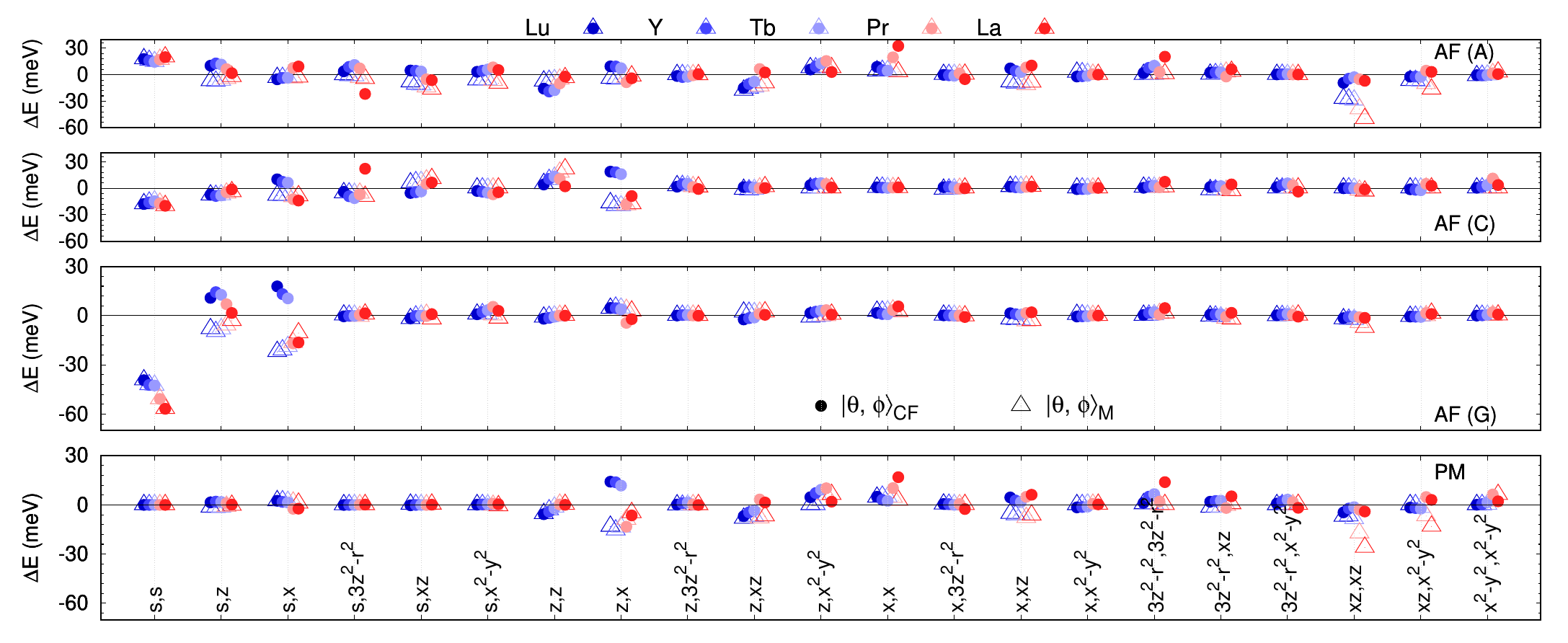}
\caption{SE energy gain $\Delta E(\theta, \phi)$  decomposed in irreducible tensor components for hole orbitals $|\theta, \phi \rangle_{\rm{CF}}$ (filled circles)  and $|\theta_{\rm M}, \phi_{\rm M} \rangle$, (empty triangles). From top to bottom: AF (A-type), AF (C-type), AF (G-type) and PM case. }\label{DESALL}
\end{figure*}
From Eq.~(\ref{SE})  we calculate $\Delta E(\theta,\phi)$, the SE energy for a given spin and orbital ordering; 
 minimizing it with respect to  the angles  yields  $|\theta, \phi \rangle_{\rm M}$.
Finally, 
in order to unravel the complexity of the emergent ordering, we resort to a novel scheme: we decompose the order parameter itself into its irreducible components, $
p_q{=}\sum_{r>0 \mu } a_{r\mu;q0} \langle \hat{\tau}_i^{r\mu; q 0} \rangle 
$; the orbital polarization is $p{=}p_0$,  while $m{=} \langle \hat{\tau}_i^{0 0; 1 0} \rangle$. 
Last, the expectation value of an operator for a specific hole orbital \cite{note} is
$\langle \hat{\tau}_i^{r,\mu} \rangle_{\rm \alpha}{=}_\alpha\langle \theta, \phi | \hat{\tau}_i^{r,\mu; q\nu} |\theta, \phi \rangle_{\rm \alpha}$.

{\em Orbital ordering in the paramagnetic phase}.
In Fig.~\ref{dmft}, panel (a), we show the orbital polarization, $p(T)$,  in the paramagnetic (PM) phase; the corresponding hole orbital 
 is $|\theta, \phi \rangle_{\rm{OO}}$. We define $T_{\rm OO}$  as the temperature for which $p(T){=}0.5$, i.e., for which
 orbital fluctuations are sizably suppressed.
The figure  shows that $T_{\rm OO}$ is appreciabily higher  for $R={\rm Lu} $ than for $R={\rm La}$.
These calculations include both  SE and lattice distortion effects.
In order to single out SE effects from the rest, we repeat the same calculations but  this time for $\varepsilon_{\rm CF}{=}{0}$, panel (b); this yields $T_{\rm KK}$, the SE transition temperature,
and the associated hole orbital, $|\theta, \phi \rangle_{\rm{KK}}$. 
The figure shows that   $T_{\rm{KK}}$ increases going from Tb to La, with a variation of at most of 60 K,
i.e., $T_{\rm{KK}}$  behaves the  {\em opposite} of $T_{\rm OO}$. 

The trend for $T_{\rm KK}$ can be understood from Fig.~\ref{DESALL}, bottom panel. It shows
the SE energy,  $\Delta E(\theta,\phi)$, split into its irreducible-tensor  components, for the ideal case $|\theta,\phi\rangle{=}|\theta, \phi\rangle_{\rm M}$, the state that maximizes the SE energy gain  (empty triangles).
For small $R_I$ the main contribution is from the  anisotropic dipolar term $(x,z)$.
Increasing $R_I$, however,  the weight shifts progressively to the  quadrupolar $(xz, xz)$ channel;
this is because the GdFeO$_3$-type distortion decreases \cite{3d1b}, which, 
ceteris paribus, reduces off-diagonal hopping integrals (see Supplemental Material \cite{SM}).
This makes SE more efficient: the maximum energy gain in the $(x,z)$ channel is obtained for states not in line with lattice symmetry \cite{symmetry}.
This explains the trends for $T_{\rm KK}$ in Fig.~\ref{dmft}: indeed, when CF effects are negligible, 
 $|\theta, \phi\rangle_{\rm KK}{\sim}|\theta, \phi\rangle_{\rm M}$.

 The trend of $T_{\rm OO}$ is more complex to unravel.
 In the real structures, CF and SE energies are comparable, and the SE energy surface alone is not sufficient.
\begin{figure*}[t]
\centering  
\includegraphics[width=0.85\textwidth]{ 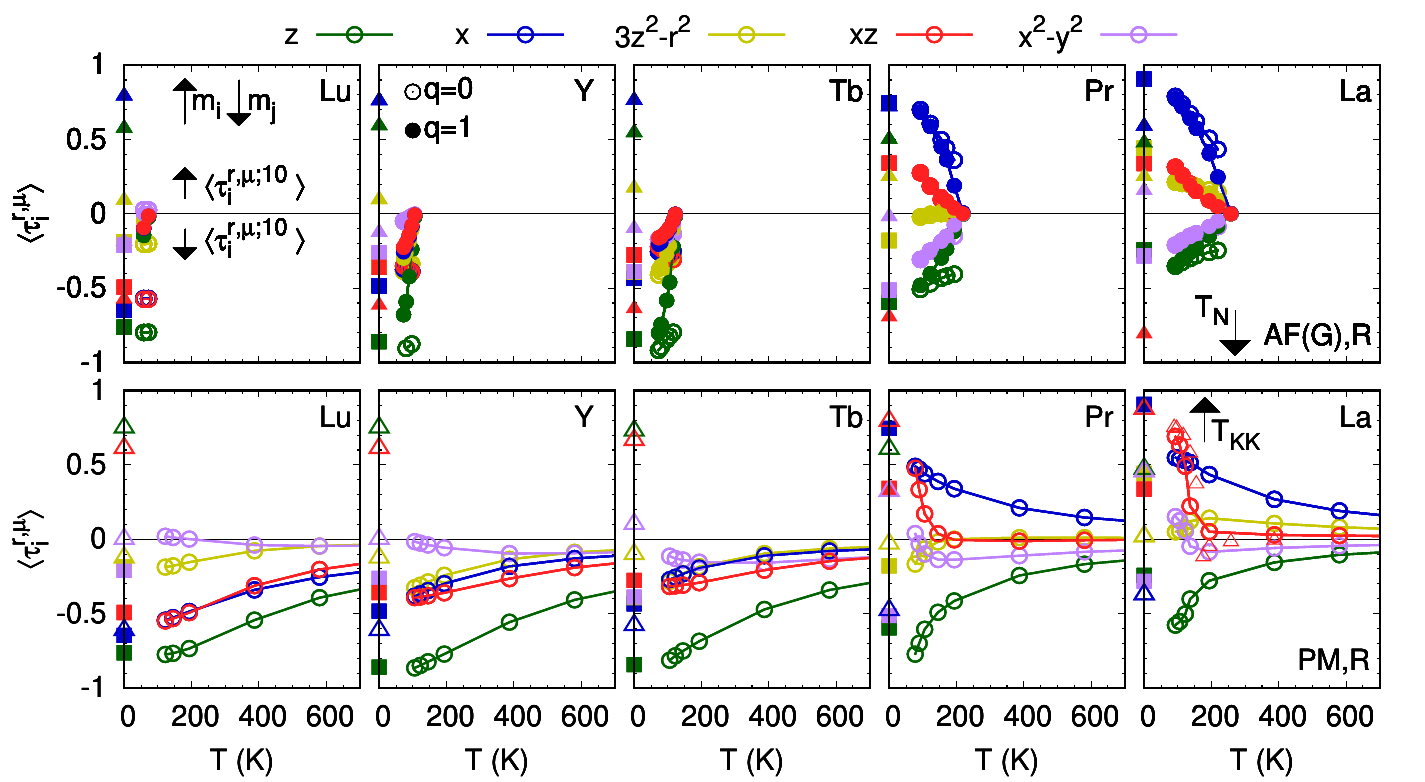}
\caption{ Irreducible components of the order parameter, $\langle \hat{\tau}_i^{r\mu;q\nu} \rangle$, normalized to the maximum value they can reach \cite{SM}. $T{=}0$ axis:
$\langle \theta, \phi | \hat{\tau}_i^{r,\mu; q\nu} |\theta, \phi \rangle_\alpha$
 for  $\alpha{=}{\rm M}$  (triangles) and $\alpha{=}{\rm CF}$ (filled square).
 Bottom: PM phase (empty symbols).
Top:  AF (G) (filled symbols). 
The magnetization is $m_i=m>0$ at site $i$  (Fig.~\ref{dmft}) and $m_j=-m$ at a neighboring site.} \label{taus}
\end{figure*}
We thus resort to a novel approach: decompose 
 the order parameter $p(T)$ itself into its irreducible components, $\langle \hat{\tau}_i^{r\mu;q\nu} \rangle$. The results  are shown in Fig.~\ref{taus}, bottom panels.
For small $R_I$  all  $\langle \hat{\tau}_i^{r\mu;00} \rangle$ rise smoothly: there is no phase transition in any channel, the signature of the presence of  SE interactions is merely in small deviations
of $\langle \hat{\tau}_i^{r\mu;00} \rangle$, the saturation moments,   from  $\langle \hat{\tau}_i^{r\mu;00} \rangle_{\rm CF}$  (filled squares). 
Increasing $R_I$, however,  a sudden change of behavior occurs: a phase transition appears in the $xz$ channel, 
very similar to the one obtained in the ideal case with no CF splitting (empty triangles) \cite{SM}.
 The
$(xz,xz)$ interaction is here the dominant orbital SE interaction as in the ideal case.
This provides the key for explaining the trends. Starting from small $R_I$, the evolution of $T_{\rm OO}$
with increasing $R_I$ essentially  reflects the CF: $T_{\rm OO}$  decreases because the CF splitting decreases. Eventually,
 $T_{\rm OO}{\sim} T_{\rm KK}$, but this happens only for large $R_I$. This explains the opposite behavior of $T_{\rm OO}$  and $T_{\rm KK}$ with  increasing $R_I$.

The resulting theoretical phase diagram for the PM phase is shown in Fig.~\ref{temp},
where the experimental structural transition $T_S$ is  also reported.
The figure shows  that $T_{\rm OO}{>}T_S$  for small $R_I$,  while  $T_{\rm OO}{\sim}T_S$ for large $R_I$.
Correspondingly, for small $R_I$, right above $T_S$ the polarization is  large ($p(T){\gtrsim}0.75$), i.e., OO is already well developed. 
Finally, $T_{\rm KK}{<}T_S$  for small $R_I$,   while $T_{\rm KK}{\sim}T_{\rm OO}{\gtrsim}T_S$ for large $R_I$.
Entering the monoclinic phase modifies the OO state, not the overall picture; it slightly enhances $T_{\rm KK}$, but the CF 
 increases as well.

{\em Antiferromagnetic phase.}
Finally, we turn to the most interesting point, the surprising inversion of $T_N$ and $T_{\rm OO}$ with increasing $R_I$.
Let us first discuss the DMFT results for the G-type magnetic phase (AF order in all directions),
Fig.~\ref{dmft}.  The figure  shows that  $T_{N}$, panel (d), increases with $R_I$, similarly to $T_{\rm KK}$, panel (b), but much more rapidly.
Thus, remarkably, our numerical calculations do yield  $T_{N}{<}T_{\rm KK}{<}T_{\rm OO}$ for small $R_I$ and $T_{N}{>}T_{\rm KK}{\sim} T_{\rm OO}$ for large $R_I$, in line with experiments. 

Next, we explain why. Our decomposition of the SE interaction in Eq.~(\ref{SE}) makes clear that the only interaction that can lead to  $T_{N}{>}T_{\rm KK}$
is the  $r{=}0$ (orbital monopolar) dipolar spin-spin term, $(s,s)$. The latter, taken alone, depends little on OO or the CF.
We find that such a term is AF in all directions, Fig.~\ref{DESALL};
in the AF (G) phase, it is also the dominant magnetic  interaction.
Its strength increases with $R_I$,  in particular along ${\bf c}$, i.e., it is weaker for small $R_I$; 
in the ideal case with no CF,  it is just strong enough to make $T_N$ slightly larger than $T_{\rm KK}$  {\em for all $R_I$}, see Fig.~\ref{dmft}, panel (e).
It is only when we switch on the CF splitting,  that  we obtain the inversion of $T_N$ and $T_{\rm KK}$ with increasing $R_I$. 

This surprising effect is a consequence of the spin-orbital coupling via the $\hat{H}_{S_iS_jO_iO_j}$ SE term.  Fig.~\ref{taus}, top panel, shows that the $q{=}1$  and  $q{=}0$ operators have similar
expectation value below $T_N$.  All the $q{=}1$ expectation values, however, alternate sign for two AF neighboring sites $(i,j)$:  thus, for $r{+}r'{>}1$, the $q{=}1$  and  $q{=}0$ contributions  tend to cancel each other, see Fig.~\ref{DESALL}, AF(G) panel, for the  limit of full magnetization. The channels influencing $T_N$ are then mostly $(s,\mu)$  terms with spin rank $q{=}1$ and $\mu=x,z$.
 For small $R_I$, in the presence of CF,   the latter are ferromagnetically  aligned  ($m_j\langle \hat{\tau}_i^{r\mu}\rangle{>}0$
for all neighbors $j$, see Fig.~\ref{taus}); this  leads to magnetic frustration, reducing $T_N$; 
for large $R_I$, instead, the alignement is antiferro in three channels, increasing $T_N$.
 This is why, for the real structures, $T_{N}{>}T_{\rm KK}{<}T_{\rm OO}$ for large $R_I$ but $T_{N}{<}T_{\rm KK}{\sim}T_{\rm OO}$  when $R_I$ is small.

Let us now consider  other possible AF magnetic structures: C-type, with ferro (F) order along {\bf c}, and A-type, with F order in the {\bf ab} plane (the latter is not found experimentally for these systems).  In both the A and C structures, for at least one bond  the ($s,s$) component of the spin-spin interaction is frustrated.
This, taken alone, reduces $T_N$.
The F stacking, either along {\bf c} or in the plane, is stabilized by  terms with orbital rank $r{+}r'{>}0$, hence by  OO, however. 
 For La, Fig.~\ref{dmft} shows that indeed $T_N{\lesssim} T_{\rm KK}$ for C-type magnetic order.
Entering the monoclinic phase, ceteris paribus, does not change this conclusion: the ($s,s)$ magnetic term
remains in fact essentially the same; C-type spin order is assisted by the CF splitting \cite{mononote}. 
Finally, in the A phase, there is no magnetic order without OO for the systems considered, since the
($s,s$) spin-spin orbital monopolar interaction  alone gives an energy loss, see  Fig.~\ref{dmft}, panel (f).

{\em Conclusion.}
In conclusion,  using a new analysis scheme based onto the decomposition of the order parameter into its irreducible components, we explained the origin of the inversion of $T_{N}$ and $T_{\rm OO}$ with increasing $R_I$  in vandates. Our results  clarify why the $T_{\rm N}$-$T_{\rm KK}$ inversion is rare: in most orbitally ordered systems
the onset of orbital ordering is determined by lattice distortions, i.e., $T_{\rm OO}$ is controlled by the CF splitting, rather than SE.
The inversion instead requires that two conditions are simultaneously verified: $(i)$ that $T_{\rm OO}{\sim}T_{\rm KK}$, and $(ii)$ that  nonetheless the magnetic structure is determined by dipolar 
interactions with orbital rank $r{=}0$. This indicates that small $R_I$ systems could be manipulated to satisfy these conditions and provides criteria to look for similar behavior in other materials. Finally, the approach used in this work should prove powerful for understanding other classes of problems, such
as the nature of octupolar order in materials with strong spin-orbit coupling.

We would like to acknowledge computational time on JURECA  as well as on JUWELS; the latter was used in particular for code development.


\begin{thebibliography}{99}
\bibitem{KB} D.I. Khomskii, {\em Transition Metal Compounds},  Cambridge University Press (2014).
{\bibitem{Sci2000} Y. Tokura and N. Nagaosa, Science {\bf 288}, 462-468 (2000).}
{\bibitem{Sci2005} E. Dagotto, Science {\bf 309}, 257-262 (2005).}

{\bibitem{Interface} H. Y. Hwang, Y. Iwasa, M. Kawasaki, B. Keimer, N. Nagaosa, and Y. Tokura, Nat. Mat. {\bf 11}, 103-113 (2012).}
\bibitem{diso}J. Yan, A. Kumar, M. Chi, M. Brahlek, T. Z. Ward, and M.A. McGuire, Phys. Rev. Mat. {\bf 8}, 024404 (2024). 
{\bibitem{nonequilibrium} A. S. Disa, J. Curtis, M. Fechner, A. Liu, A. von Hoegen, M. F\"orst, T. F. Nova, P. Narang, A. Maljuk, A. V. Boris, B. Keimer, and A. Cavalleri, Nature {\bf 617}, 73-78 (2023).}


\bibitem{Khaliullin2007} P. Horsch, A. M. Ole\'s, L. F. Feiner and G. Khaliullin, Phys. Rev. Lett. {\bf 100}, 167205 (2008).
{\bibitem{structure} M. J. Mart\'inez-Lope, J. A. Alonso, M. Retuerto, and M. T. Fern\'andez-Díaz, Inorg. Chem. {\bf 47}, 2634 (2008).}

{\bibitem{Goodenough2008} J.-S. Zhou, Y. Ren, J.-Q. Yan, J.F. Mitchell, and J. B. Goodenough, Phys. Rev. Lett. {\bf 100}, 046401 (2008).}
\bibitem{Ren2003} Y. Ren, A. A. Nugroho, A. A. Menovsky, J. Strempfer, U. R\"utt, F. Iga, T. Takabatake and C. W. Kimball, Phys. Rev. B {\bf 67}, 014107 (2003).
\bibitem{Goodenough2004} J.-Q. Yan, J.-S. Zhou, and J. B. Goodenough, Phys. Rev. Lett. {\bf 93}, 235901 (2004).
{\bibitem{Palstra2006} M. H. Sage, G. R. Blake, G. J. Nieuwenhuys, and T. T. M. Palstra, Phys. Rev. Lett. {\bf 96}, 036401 (2006).}
\bibitem{Sund2023}P. N. Shanbhag, F.  Fauth, and A. Sundaresan, Phys. Rev. B {\bf 108} 134115 (2023).
\bibitem{Nasir2023}M. Nasir, I. Kim, Ki. Lee, S. Kim, K. Hyoung Lee, and H. Jung Park
Phys. Chem. Chem. Phys. {\bf 25}, 3942 (2023).
\bibitem{Meley2021} H. Meley, M. Tran, J. Teyssier, J. A. Krieger, T. Prokscha, A. Suter, Z. Salman, M. Viret, D. van der Marel, and S. Gariglio,
Phys. Rev. B {\bf 103}, 125112 (2021).
\bibitem{Yan2019}
J.-Q. Yan, W. Tian, H. B. Cao, S. Chi, F. Ye, A. Llobet, A. Puretzky, Q. Chen, J. Ma, Y. Ren, J.-G. Cheng, J.-S. Zhou, M. A. McGuire, and R. J. McQueeney,
Phys. Rev. B {\bf 100} 184423 (2019).
\bibitem{Gruninger2013}
E. Benckiser, L. Fels, G. Ghiringhelli, M. Moretti Sala, T. Schmitt, J. Schlappa, V. N. Strocov, N. Mufti, G. R. Blake, A. A. Nugroho, T. T. M. Palstra, M. W. Haverkort, K. Wohlfeld, and M. Gr\"uninger,
Phys. Rev. B, {\bf 88} 205115 (2013).
\bibitem{Parmigiani2012} 
F. Novelli, D. Fausti, J. Reul, F. Cilento, P. H. M. van Loosdrecht, A. A. Nugroho, T. T. M. Palstra, M. Gr\"uninger, and F. Parmigiani,
Phys. Rev. B {\bf 86}, 165135 (2012).
\bibitem{Gatti2021} 
K. Ruotsalainen, M. Gatti, J.M. Ablett, F. Yakhou-Harris, J.-P. Rueff, A. David, W. Prellier, and A. Nicolaou,
Phys. Rev. B {\bf 103}, 235158 (2021).
\bibitem{Palstra2007} M. H. Sage, G. R. Blake, C. Marquina, and T. T. M. Palstra, Phys. Rev. B {\bf 76}, 195102 (2007).
{\bibitem{Palstra2001} G. R. Blake, T. T. M. Palstra, Y. Ren, A. A. Nugroho, and A. A. Menovsky, Phys. Rev. Lett. {\bf 87}, 245501 (2001).}
{\bibitem{Keimer2011} M. Reehuis, C. Ulrich, K. Proke\v{s}, S. Mat\'a\v{s}, J. Fujioka, S. Miyasaka, Y. Tokura, and B. Keimer, Phys. Rev. B {\bf 83}, 064404 (2011).}
{\bibitem{Sundaresan2017} R. Saha, F. Fauth, V. Caignaert, and A. Sundaresan, Phys. Rev. B {\bf 95}, 184107 (2017).}
\bibitem{Keimer2008} M. Reehuis, C. Ulrich, P. Pattison, M. Miyasaka, Y. Tokura, and B. Keimer, Eur. Phys. J. B {\bf 64}, 27–34 (2008).
\bibitem{Tokura2003} S. Miyasaka, Y. Okimoto, M. Iwama, and Y. Tokura, Phys. Rev. B {\bf 68}, 100406(R) (2003).

{\bibitem{Ren2007} J.-S. Zhou, J. B. Goodenough, J.-Q. Yan, and Y. Ren, Phys. Rev. Lett. {\bf 99}, 156401 (2007).}
{\bibitem{Fernandez-Diaz2003} A. Munoz, J. A. Alonso, M. T. Cas\'ais, M. J. Mart\'inez-Lope, J. L. Mart\'inez, and M. T. Fern\'andez-D\'iaz, Phys. Rev. B {\bf 68}, 144429 (2003).}
\bibitem{Reul} J. Reul, A. A. Nugroho, T. T. M. Palstra, and M. Gr\"uninger, Phys. Rev. B {\bf 86}, 125128 (2012).
\bibitem{Molly2007} M. De Raychaudhury, E. Pavarini, and O. K. Andersen, Phys. Rev. Lett. {\bf 99}, 126402 (2007).
\bibitem{xj2022} X.-J. Zhang, E. Koch, and E. Pavarini, Phys. Rev. B {\bf 105}, 115104 (2022).


\bibitem{KK} K.I. Kugel and D.I. Khomskii, Zh. Eksp. Teor. Fiz. {\bf 64}, 1429 (1973) [Sov. Phys. JETP {\bf 37}, 725 (1973)].
\bibitem{2xj2022} X.-J. Zhang, E. Koch, and E. Pavarini, Phys. Rev. B {\bf 106}, 115110 (2022).

\bibitem{oo1}E. Pavarini, E. Koch, A.I. Lichtenstein, Phys. Rev. Lett. {\bf 101}, 266405 (2008).
\bibitem{oo2}E. Pavarini and E. Koch, Phys. Rev. Lett. {\bf 104}, 086402 (2010).
\bibitem{xj2020} X.-J. Zhang, E. Koch, and E. Pavarini, Phys. Rev. B {\bf 102}, 035113 (2020).
\bibitem{oo4} J. Musshoff, G. Zhang, E. Koch, E. Pavarini, Phys. Rev. B {\bf 100}, 045116 (2019).
\bibitem{oo5} C. Autieri, E. Koch, and E. Pavarini, Phys. Rev. B {\bf 89}, 155109 (2014).

\bibitem{BLAHA1990399} P. Blaha, K. Schwarz, P. Sorantin, and S. Trickey, Comput. Phys. Commum. {\bf 59}, 399 (1990).
\bibitem{PhysRevB.56.12847} N. Marzari and D. Vanderbilt, Phys. Rev. B {\bf 56}, 12847 (1997).
\bibitem{wannier90} A. A. Mostofi, J. R. Yates, Y.-S. Lee, I. Souza, D. Vanderbilt, and N. Marzari, Comput. Phys. Commun. {\bf 178}, 685 (2008); J. Kune\'s, R. Arita, P. Wissgott, A. Toschi, H. Ikeda, and K. Held, {\emph {ibid.}} {\bf 181}, 1888 (2010).
\bibitem{PhysRevB54.5368} T. Mizokawa and A. Fujimori, Phys. Rev. B {\bf 54}, 5368 (1996).
\bibitem{3d1b} E. Pavarini, A. Yamasaki, J. Nuss, and O. K. Andersen, New J. Phys. {\bf 7}, 188 (2005).
\bibitem{RevModPhys.83.349} E. Gull, A. J. Millis, A. I. Lichtenstein, A. N. Rubtsov, M. Troyer, and P. Werner, Rev. Mod. Phys. {\bf 83}, 349 (2011).
\bibitem{lamno3c} A. Flesch, E. Gorelov, E. Koch, and E. Pavarini, Phys. Rev. B {\bf 87}, 195141 (2013).
\bibitem{julian1} J. Musshoff, G. R. Zhang, E. Koch, and E. Pavarini, Phys. Rev. B {\bf 100}, 045116 (2019).
\bibitem{julian2} J. Musshoff, A. Kiani, and E. Pavarini,  Phys. Rev. B {\bf 103}, 075136 (2021).

\bibitem{symmetry} Defining $V_1$ and $V_2$ two neighboring sites in one layer and $V_3$ and $V_4$ the corresponding ones in the layer above, with $V_3$ on top of $V_1$,  in the GdFeO$_3$-type structure, if  $|\theta,\phi\rangle^1$ is the hole orbital at site $V_1$, the  hole orbitals at the other sites are $|\theta,\phi\rangle^2{=}|\theta,\frac{\pi}{2}{-}\phi\rangle^1$, $|\theta,\phi\rangle^3{=}|-\theta,\phi\rangle^1$, and $|\theta,\phi\rangle^4{=}|-\theta,\frac{\pi}{2}-\phi\rangle^1$, respectively.  In the monoclinic case, the relation $|\theta,\phi\rangle^3{\sim}|-\theta{-}\delta_\theta,\phi{\pm}\delta_\phi\rangle$ maximizes the energy gain  with $\delta_\theta{\sim}20^{\circ}$ and $\delta_\phi{\sim}110^{\circ}$ in the PM phase.
 \bibitem{SM} For additional information see Supplemental Material.


\bibitem{note} For a given spin rank $q$, there are only two independent  saturation ($T{\to}0$) values.
This is because, for a specific hole orbital, 
$\langle \hat{\tau}_i^{r,\mu} \rangle_{\rm \alpha}{=}\langle \theta, \phi | \hat{\tau}_i^{r,\mu; q\nu} |\theta, \phi \rangle_{\rm \alpha}$, 
 depend on two  variables only,  $\theta$ and $\phi$  \cite{SM}. These constraints do not  hold at finite temperature, however. 
 
 \bibitem{mononote} The CF affects $T_N$ more for C-type than G-type magnetic order, making the first more stable. 

 
 
\end{thebibliography}
\end{document}